\documentstyle[aps,preprint,tighten,epsfig]{revtex}
\def\beq{\begin{equation}}
\def\eeq{\end{equation}}
\def\beqa{\begin{eqnarray}}
\def\eeqa{\end{eqnarray}}
\def\MeV{\nobreak\,\mbox{MeV}}
\def\GeV{\nobreak\,\mbox{GeV}}

\def\pli{p^\prime}
\def\mli{{M^\prime}^2}
\begin{document}
\title{\sc  $D^*D\pi$ form factor revisited}
\author {F.S. Navarra, M. Nielsen \\
\vspace{0.3cm}
{\it Instituto de F\'{\i}sica, Universidade de S\~{a}o Paulo, } \\
{\it C.P. 66318,  05389-970 S\~{a}o Paulo, SP, Brazil}\\
\vspace{0.3cm}
M.E. Bracco\\
\vspace{0.3cm}
{\it Instituto de F\'{\i}sica,  Universidade do Estado do Rio de Janeiro, } \\ 
{\it Rua S\~ao Francisco Xavier 524, Maracan\~a,  20559-900, Rio de Janeiro, 
RJ, Brazil}}
\maketitle
\vspace{1cm}

\begin{abstract}
The $D^*D\pi$ form factor  is evaluated in a QCD sum rule calculation 
for both $D$ and $\pi$ off-shell mesons. 
We study the Borel sum rule for the three
point function of one pseudoscalar, one axial and one vector meson currents. 
We find that the momentum dependence of the form factors is very different
if the $D$ or the $\pi$ meson is off-shell, but they lead to the same 
coupling constant in the $D^*D\pi$ vertex. 
\\
PACS numbers 14.40.Lb,~~14.40.Nd,~~12.38.Lg,~~11.55.Hx
\\

\end{abstract}

\vspace{1cm}
	In a very recent measurement by the CLEO collaboration \cite{cleo},
the total width of $D^*$ meson was obtained: $\Gamma_{tot}(D^*)= 96\pm4\pm22$
keV. This measurement yields the strong $D^*D\pi$ coupling, $g_{D^*D\pi}=
17.9\pm0.3\pm1.9$, which is defined as
\cite{bel}
\beq
\langle D^{*-}(p)|\pi^-(p') D^0(p-p')\rangle=g_{D^*D\pi}p'_\mu\epsilon^\mu\;.
\label{con}
\eeq

The $D^*D\pi$  coupling constant has been studied by several authors 
using different approaches of the QCD sum rules (QCDSR): two point function
combined with soft pion techniques \cite{col,ele}, light cone sum rules
\cite{bel,col2}, light cone sum rules including perturbative corrections 
\cite{kho}, sum rules in a external field \cite{gro}, double
momentum sum rules \cite{dn}, double Borel sum rules \cite{ddpi}. 
Unfortunately, the numerical results from these calculations may differ by 
almost a factor two, and the upper limit of these predictions give
$g_{D^*D\pi}=13.5$ \cite{kho}, which is still 25\% lower than the central 
value of the CLEO measurement.

In ref.~\cite{ddpi} we have estimated the $g_{D^*D\pi}(Q^2)$ form factor as
a function of the off-shell pion momentum $Q^2$. Since the sum rule obtained
from the used three-point function is not valid at $Q^2=0$, in order to 
determine
the $D^*D\pi$ coupling we had to extrapolate the $Q^2$ behaviour of
the form factor. Of course there are large uncertainties in this procedure, 
and,
to guide us in choosing the analytical form which parametrizes our QCDSR
results, we have used a QCDSR valid only at $Q^2=0$, as suggested in 
\cite{rry} for the pion-nucleon coupling constant. We got \cite{ddpi}
$g_{D^*D\pi}=5.7\pm 0.4$, a value much smaller than the experimental result.
However, in a recent calculation of the $DD\rho$ form factor \cite{ddrho}, we
have used a completely different approach to get the $DD\rho$ coupling 
constant: we have calculated the $DD\rho$ form factor for both $D$ and $\rho$
off-shell mesons, and the QCDSR results were parametrized by analytical forms
such that the respective extrapolations at the $D$ and $\rho$ poles
provided consistent values for the $g_{DD\rho}$ coupling constant.

In this work we use the three-point function approach to evaluate the 
$D^*D\pi$  form factor for a off-shell $D$ meson, and we will follow the
procedure suggested in ref.~\cite{ddrho} to parametrize the $Q^2$ behaviour
of the $D^*D\pi$ form factor for both $D$ and $\pi$ off-shell mesons.

The three-point function associated with a $D^*D\pi$ vertex  with an
off-shell $D$ meson is given by
\begin{equation}
\Gamma_{\mu\nu}(p,\pli)=\int d^4x \, d^4y \, \langle 0|T\{j_\nu(x)
j_D(y)j^\dagger_\mu(0)\}|0\rangle  
\, e^{ip^\prime.x} \, e^{i(p-\pli).y}\; , 
\label{cor}
\end{equation}
where $j_D=i\bar{c}\gamma_5 u$, $j_\nu=\bar{u}\gamma_\nu \gamma_5d $ and
$ j_\mu=\bar{c}\gamma_\mu d$ are the interpolating fields for $D^0$, 
$\pi^-$ and $D^{*-}$ respectively with $u$, $d$ and $c$ being the up, down, 
and charm quark fields.

The phenomenological side of the vertex function, $\Gamma_{\mu\nu}
(p,p^\prime)$,
is obtained by the consideration of $\pi$ and $D^*$ state contribution to
the matrix element in Eq.~(\ref{cor}):

\beqa
\Gamma_{\mu\nu}^{(phen)}(p,\pli)&=&i{f_\pi f_D f_{D^*} m_{D^*}(m_D^2/m_c)
g_{D^*D\pi}(q^2)\over (q^2-m_D^2)
(p^2-m_{D^*}^2)({\pli}^2-m_\pi^2)}\times
\nonumber \\*[7.2pt]
&&\left(\pli_\mu\pli_\nu+{m_{D^*}^2+m_\pi^2-q^2\over2m_{D^*}^2}p_\mu\pli_\nu
\right)
+ \mbox{higher resonances}\; .
\label{phen}
\eeqa

To derive Eq.~(\ref{phen}) we have made use of the generalization of 
Eq.~(\ref{con}) for an off-shell $D$ meson: $\langle D^{*-}(p)|\pi^-(p') 
D^0(q)\rangle=g_{D^*D\pi}(q^2)p'_\mu\epsilon^\mu$, where $q=p-\pli$, and the 
decay constants $f_\pi,\;f_D$ and $f_{D^*}$ defined by the matrix elements
\beq
\langle 0|j_\nu|\pi(\pli)\rangle=if_\pi\pli_\nu\;,
\label{fpi}
\eeq
\beq
\langle 0|j_D|D\rangle={m_D^2f_D\over m_c}\;,
\label{fh}
\eeq
and
\beq
\langle D^*|j^\dagger_\mu|0\rangle=m_{D^*}f_{D^*}\epsilon^*_\mu
\; ,
\label{fd*}
\eeq
where $\epsilon^\nu$ is the polarization of the vector meson. 
The contribution of higher resonances and continuum in Eq.~(\ref{phen})
will be taken into account as usual in the standard form of 
ref.~\cite{io2}, through the continuun thresholds $s_0$ and $u_0$, for the 
$D^*$ and $\pi$ mesons respectively.

The QCD side, or theoretical side, of the vertex function is evaluated by
performing Wilson's operator product expansion (OPE) of the operator
in Eq.~(\ref{cor}). Writing $\Gamma_{\mu\nu}$ in terms of the invariant
amplitudes,
we can write a double dispersion relation for each one of the invariant
amplitudes, over the virtualities $p^2$ and ${\pli}^2$
holding $Q^2=-q^2$ fixed:
\beq
\Gamma(p^2,{\pli}^2,Q^2)=-{1\over4\pi^2}\int_{m_Q^2}^{s_0} ds
\int_0^{u_0} du {\rho(s,u,Q^2)\over(s-p^2)(u-{\pli}^2)}\;,
\label{dis}
\eeq
where $\rho(s,u,Q^2)$ equals the double discontinuity of the amplitude
$\Gamma(p^2,{\pli}^2,Q^2)$ on the cuts $m_Q^2\leq s\leq\infty$,
$0\leq u\leq\infty$, which can be evaluated using Cutkosky's rules
\cite{io2,cut}. 
Finally we perform a double Borel transformation \cite{io2} in both variables
$P^2=-p^2\rightarrow M^2$ and ${P^\prime}^2=-{\pli}^2\rightarrow \mli$ and 
equate the two representations
described above. We get one sum rule for each invariant function. In the
$\pli_\mu\pli_\nu$ structure the double discontinuity of the perturbative 
contribution reads:
\beqa
\rho(s,u,t)&=&-{6im_c\over(\lambda(s,u,t))^{5/2}}\left[m_c^4(\lambda(s,u,t)
+6su)+s\left(
s^2(t+u)+(t-u)^2(t+u)-2s(t^2\right.\right.\nonumber\\
&-&\left.\left.tu+u^2)\right)+m_c^2\left(-s^3+s^2(t-3u)-(t-u)^3+s(t^2-
4tu+3u^2)\right)\right]
\; ,
\eeqa
where $t=-Q^2$ and $\lambda(s,u,t)=s^2+u^2+t^2-2su-2st-2tu$.
The integration limit condition is
\beq
u\leq s+t-m_c^2-{st\over m_c^2}\; .
\eeq

For consistency we use in our analysis the QCDSR 
expressions for the $D^*$ and $\pi$ decay constants up to dimension four 
in lowest order  of $\alpha_s$ as given in refs.~\cite{bel,ddpi,svz}.

The parameter values used in all calculations are 
$m_c=1.5\,\GeV$, $m_\pi=140\,\MeV$, $m_D=1.87\,\GeV$, $m_{D^*}=2.01\,\GeV$, 
$f_D=160\,\MeV$,
$\langle\overline{q}q\rangle\,=\,-(0.23)^3\,\GeV^3$, $\langle g^2G^2\rangle 
=0.5\,\GeV^4$, $s_0=6.3\,\GeV^2$ and $u_0=2.0\,\GeV^2$.

In the calculation of the $D^*D\pi$ form factor with the off-shell pion 
\cite{ddpi} we have included, besides the perturbative contribution, the
gluon condensate contribution. We have found out that the gluon condensate
is small, as compared with the perturbative contribution and decreases
with the Borel mass. The most important feature of the gluon condensate is 
the fact that it improves the stability of the result as a function of the 
Borel mass. Since its contribution at $M^2=7\GeV^2$, is less than 5\% 
of the perturbative contribution,
in this work we will neglect the gluon condensate. In order  to be sure  
that the absence of the  gluon condensate will not affect our results, 
we will extract the value of the
form factor at a higher value of the Borel mass, where we expect  the
gluon condensate contribution to be negligible. 

In refs.~\cite{BBG93,ra} it was found that relating 
the Borel parameters in the two- $(M_M^2)$ and three-point functions $(M^2)$
as
\begin{equation}
         2M_{M}^2=M^2 ~ ,
\label{borel}
\end{equation}
is a crucial
ingredient for the incorporation of the HQET symmetries, and leads
to a considerable reduction of the sensitivity to input parameters, 
such as continuum thresholds $s_0$ and $u_0$, and to radiative 
corrections. Therefore, in this work we will use Eq.~(\ref{borel}) to relate
the Borel masses.

Fixing $M^2=7\,\GeV^2$ (at a fixed ratio $\mli/M^2=m_\rho^2/(m_{D^*}^2-m_c^2)$
which corresponds to $\mli=2.5\,\GeV^2$ ) we show, in 
Fig.~1,
the momentum dependence of the form factor (circles for an off-shell $D$ meson)
in the interval $-0.5\leq Q^2\leq 5\GeV$, where we expect the sum rules to be 
valid (since in this case the cut in the $t$ channel starts at 
$t\sim m_c^2$ and thus the Euclidian region stretches up to that threshold).
From this figure we can see that the $Q^2$ dependence of the 
form factor represented by the circles  can be
well reproduced by the monopole parametrization (solid line)
\beq
g_{D^*D\pi}^{(D)}(Q^2)= {126.1\over Q^2+11.95}\;.
\label{mo}
\eeq

In Fig.~1 we also show, through the squares, the momentum dependence of the 
$g_{D^*D\pi}^{(\pi)}(Q^2)$ form factor for a off-shell pion, obtained in 
ref.~\cite{ddpi}, in the interval $2\leq Q^2\leq 5\GeV$. In ref.~\cite{ddpi} 
the $Q^2$ dependence of the form factor, represented by the squares,  was
parametrized by a gaussian form (dashed line)
\beq
g_{D^*D\pi}^{(\pi)}(Q^2)=5.7 \,  e^{-Q^4/9.17}\,.
\label{ga}
\eeq
However, as can be seen by the dot-dashed line, the $Q^2$ dependence of the
QCDSR results for $g_{D^*D\pi}^{(\pi)}(Q^2)$ can also be well reproduced
by the exponential parametrization 
\beq
g_{D^*D\pi}^{(\pi)}(Q^2)=15.5 \,  e^{-Q^2/1.48}\,.
\label{exp}
\eeq
Off course, the two parametrizations in Eqs.~(\ref{ga}) and (\ref{exp}) lead
to very different values for the $D^*D\pi$ coupling constant, defined as
the value of the form factor at the pole of the off-shell meson ($Q^2=-m_\pi^2
\sim0$ in the case of the off-shell pion):
\beq
g_{D^*D\pi}=\left\{\begin{array}{l}
            5.7\;\;\;\;\;\mbox{with the gaussian parametrization}\\
            15.5\;\;\;\;\;\mbox{with the exponential parametrization}\\
           \end{array}\right.
\eeq

As discussed in the introduction, the parametrization of Eq.~(\ref{ga}),
adopted in ref.~\cite{ddpi}, was oriented by the QCDSR valid only at $Q^2=0$, 
as suggested in \cite{rry} for the pion-nucleon coupling constant. It 
consists in neglecting the pion mass in the denominator of the phenomenological
side and working at $Q^2=0$, making a single Borel 
transformation to both external momenta $P^2={P^\prime}^2\rightarrow M^2$.
The problem of doing a single Borel transformation in a three-point function
is the fact that the single pole contribution, associated
with the pole-continuum transitions, is not suppressed \cite{col,bel,io1}.
In ref.~\cite{io1} it was explicitly shown that the
pole-continuum transition has a different behavior as a function of the
Borel mass, as compared with the double pole contribution and continuum
contribution: it grows with $M^2$ as compared with the double pole 
contribution. Therefore, the single pole contribution can be taken into 
account through the introduction of a parameter $A$, in the phenomenological 
side of the sum rule \cite{bel,io1}. The value of the 
coupling constant is obtained by the extrapolation of the line $g_{D^*D\pi}
+AM^2$ to $M^2=0$ \cite{io1}. Off course this procedure also involves large 
uncertainties if $A$ is not much smaller than 1, which was the case. Also,
if $A\sim1$, this may be an indication that the sum rule is dominated
by the pole-continuum transitions and, therefore, is not a good sum rule to 
extract informations about the low-energy states.

From the parametrization in Eq.~(\ref{mo}) we can also extract the
$D^*D\pi$ coupling constant, which now is defined as
the value of the form factor at the $D$ pole ($Q^2=-m_D^2$). We get
\beq
g_{D^*D\pi}=14.9\;,
\eeq
in an excelent agreement with the exponential parametrization of
$g_{D^*D\pi}^{(\pi)}(Q^2)$. 

There is another important information that we can extract form the 
parametrization of the QCDSR results which is the value of the cut-off.
Defining  the coupling constant as the value of the form factor at 
$Q^2=-m_M^2$, where $m_M$ is the 
mass of the off-shell meson, the monopole and the exponential parametrizations
of the form factor can be written as (neglecting $m_\pi^2$):
\beq
g_{D^*D\pi}^{(D)}(Q^2)= g_{D^*D\pi}{\Lambda^2_D-m_D^2\over Q^2+\Lambda^2_D}\;,
\label{mom}
\eeq
\beq
g_{D^*D\pi}^{(\pi)}(Q^2)= g_{D^*D\pi}e^{-{Q^2\over 
\Lambda^2_\pi}}\;,
\label{expm}
\eeq
and from Eqs.~(\ref{mo}) and (\ref{exp}) we get
\beq
\Lambda_D\,=\,3.5\,\GeV\,,
\label{lad}
\eeq
\beq
\Lambda_\pi\,=\,1.2\,\GeV\,.
\label{lapi}
\eeq

Therefore, the form factor is harder if the 
off-shell meson is heavy, implying that the size of the vertex depends
on the exchanged meson, in agreement with our findings in 
refs.~\cite{ddrho,dnn}. This means that
a heavy meson will see the vertex as pointlike, whereas a 
light meson will see its extension. The value obtained for the cut-offs
are also in a very good agreement with the values of the cut-offs in the 
$DD\rho$ vertex \cite{ddrho}.

The same calculation can be done for the $B^*B\pi$ form factor and one has only
to change the $D^*,\,D$ and quark $c$ masses by $B^*,\,B$ and quark $b$ masses,
that we take as: $m_{B^*}=5.33\,\GeV$, $m_B=5.28\,\GeV$ and $m_b=4.7\,\GeV$.
In Fig.~2 we show, through the circles and through the squares, the QCDSR
results for the $B^*B\pi$ form factor with the $B$ and $\pi$ off-shell
mesons respectively. Using Eqs.~(\ref{mom}) and (\ref{expm}) to fit ours 
QCDSR results we get the couplings and cut-offs shown in Table I.

\vskip 5mm
\begin{center}
\begin{tabular}{|c|c|c|}
\hline
&&\\
& $g_{B^*B\pi}^{(M)}(Q^2=-m_M^2)$ & $\Lambda_M\,(\GeV)$\\
\hline\hline
 $B$ off-shell & 42.3& 6.8 \\
 $\pi$ off-shell & 45.1& 1.3 \\
\hline 
\end{tabular}
\end{center}
\begin{center}
\bf{TABLE I:} {\small Values of the coupling constants and cut-offs
which reproduce the QCDSR results for $g_{B^*B\pi}^{(M)}(Q^2)$.}
\end{center}
\vskip5mm

In Fig.~2 we also show, for completeness, the gaussian fit obtained in 
ref.~\cite{ddpi}, which leads to a much smaller value to the coupling constant.
It is interesting to notice that the value of the cut-off for a off-shell pion
is of the same order in both, $B^*B\pi$ and $D^*D\pi$, vertices. However, 
in the case of an off-shell $B$ meson, the cut-off is much bigger, as expected
from the discussion above.

From $g_{B^*B\pi}$ we can extract the effective scale-independent coupling
constant $g$, which controls the interaction of the pion with
infinitely heavy fields in effective lagrangian approaches \cite{cas,sin}, 
defined as $
g={f_\pi\over2m_B} g_{B^*B\pi}$.
During the last years, a large number of theoretical papers has 
been devoted to the calculation of $g$. However, the variation of the
value obtained for $g$, even within a single class of models, turns out
to be quite large. For instance, using different quark models one obtains
$1/3\leq g\leq 1$ \cite{sin} while  QCDSR calculations points in the
direction of small $g$, with a typical value in the range
$g\simeq0.13 - 0.35$ \cite{bel,col,ele,col2,gro,dn}.

Using the values for $g_{B^*B\pi}$ given in Table I we get, at order 
$\alpha_s=0$: 
\beq
g=0.59 - 0.63\;,
\eeq
therefore, our number is much bigger than the other
QCDSR calculations, and is in a better agreement with quark models.

In conclusion, we have extracted the $D^*D\pi$ coupling constant using two 
different
QCDSR for the $D^*D\pi$ form factor for the $D$ and the $\pi$ off-shell 
mesons. We have obtained for the coupling constant:
\beq
g_{D^*D\pi}=14.0\pm1.5\;,
\label{gd}
\eeq
where the errors reflect variations in the continuum thresholds,
different parametrizations of the form factors and the use of different 
relations between the Borel masses in the two- and three-point functions. 
There are still sources of errors in the values of the 
condensates and in the choice of the Borel mass to extract the form factor, 
which were not considered here. Therefore, the errors quoted are probably 
underestimated. As for the form factors,
we obtain a harder (softer) form factor when the off-shell particle is 
heavier (lighter).

In Table II we present a compilation of the estimates of the coupling
constants $g_{D^*D\pi}$ and $g_{B^*B\pi}$  from distinct QCDSR 
calculations.
\vskip 5mm
\begin{center}
\begin{tabular}{|c|c|c|}
\hline
approach & $g_{D^*D\pi}$ &  $g_{B^*B\pi}$ \\
\hline\hline
this work & $14.0\pm 1.5$ & $42.5\pm2.6$\\
two-point function + soft pion techniques (2PFSP)\cite{col} & $9\pm2$ & 
$20\pm4$\\
2PFSP + perturbative corrections \cite{col} & $7\pm2$ & $15\pm4$\\ 
light cone sum rules (LCSR) \cite{bel} & $11\pm2$ & $28\pm6$ \\
LCSR + perturbative corrections \cite{kho} & $10.5\pm3$ & $22\pm9$\\
double momentum sum rule \cite{dn} & $6.3\pm1.9$ & $14\pm4$\\ 
\hline 
\end{tabular}
\end{center}
\begin{center}
\bf{TABLE II:} {\small Summary of QCDSR estimates for  $g_{D^*D\pi}$ and
 $g_{B^*B\pi}$.}
\end{center}
\vskip5mm

From this Table we see that our result is in a fair agreement with the
LCSR calculation in refs.~\cite{bel,kho}, but is still smaller than the
experimental value \cite{cleo}: $g_{D^*D\pi}=17.9\pm0.3\pm1.9$.

\vspace{1cm}
 
\underline{Acknowledgements}: 
This work has been supported by CNPq and FAPESP.

\begin{figure} \label{fig1}
\begin{center}
\epsfysize=8.0cm
\epsffile{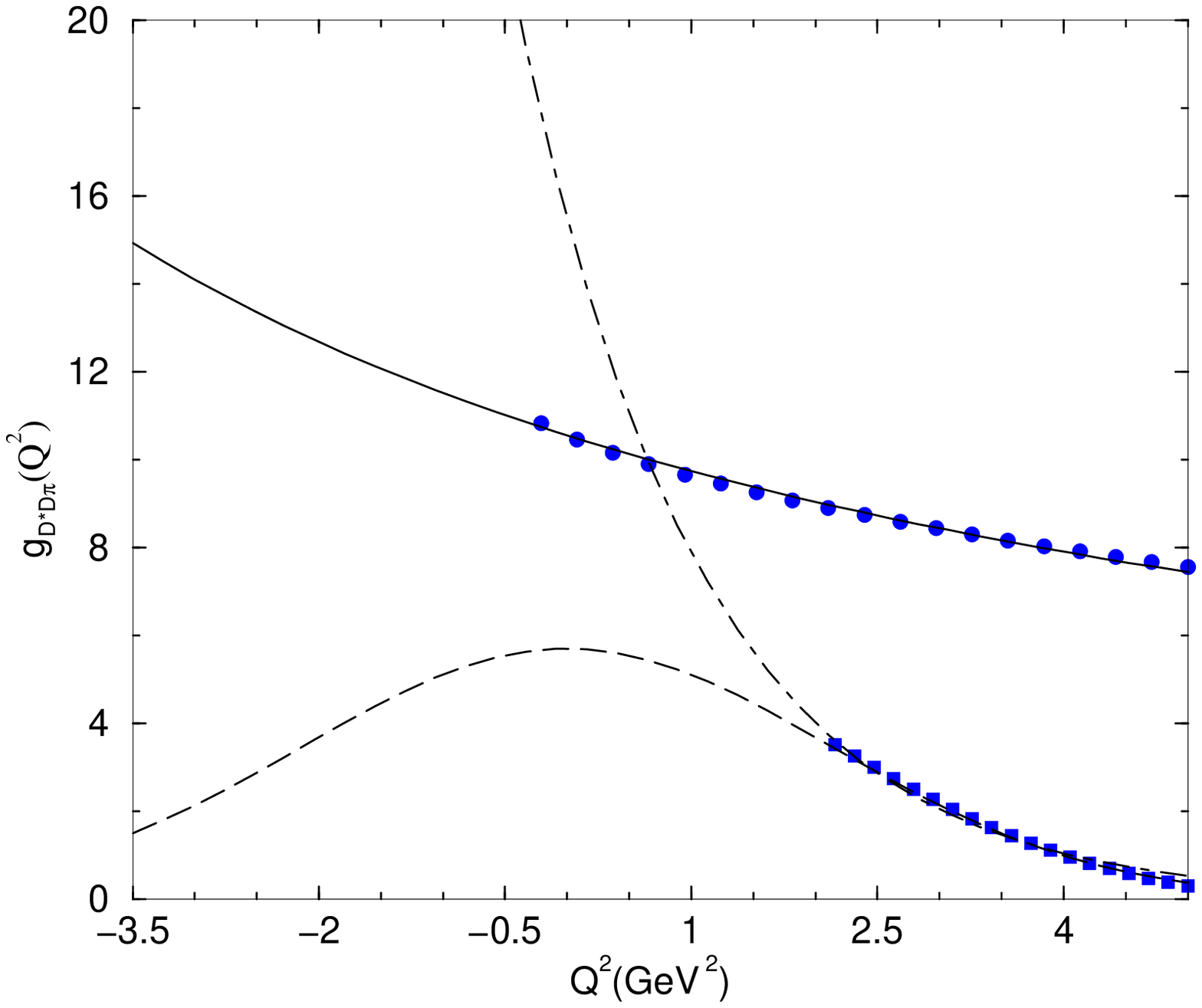}
\caption{Momentum dependence of the $D^*D\pi$ form factor.
The solid, dashed and dot-dashed lines give the 
parametrization of the QCDSR results through Eq.~(\protect\ref{mo}) for the 
circles, and Eqs.~(\protect\ref{ga}) and (\protect\ref{exp}) for the squares.}
\end{center}
\end{figure}

\begin{figure} \label{fig2}
\begin{center}
\epsfysize=8.0cm
\epsffile{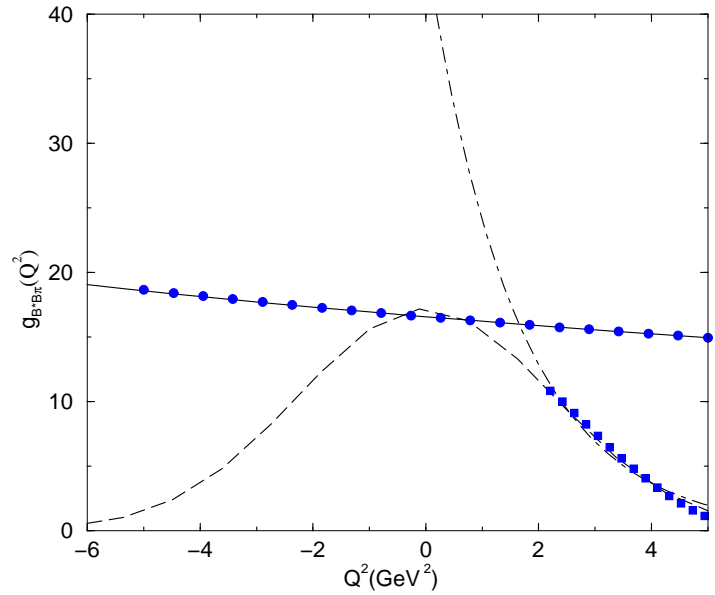}
\caption{Momentum dependence of the $B^*B\pi$ form factor.
The solid, dashed and dot-dashed lines give the 
parametrization of the QCDSR results through Eq.~(\protect\ref{mo}) for the 
circles, and Eqs.~(\protect\ref{ga}) and (\protect\ref{exp}) for the squares.}
\end{center}
\end{figure}


\vspace{0.5cm}


\begin{thebibliography}{99}
\bibitem{cleo} S. Ahmed et al., (CLEO Collaboration), hep-ex/0108013.

\bibitem{bel} V.M. Belyaev et al., {\sl Phys. Rev.} {\bf D51}, 6177 (1995). 

\bibitem{col} P. Colangelo et al., {\sl Phys. Lett.} {\bf B339}, 151 (1994).

\bibitem{ele} V.L. Eletsky and Ya.I. Kogan, {\sl Z. Phys.}  {\bf C28}, 155 
(1985);
A.A. Ovchinnikov, {\sl Sov. J. Nucl. Phys.} {\bf50}, 519 (1989).

\bibitem{col2} P. Colangelo and F. De Fazio, {\sl Eur. Phys. J.} {\bf C4},
503 (1998).

\bibitem{kho} A. Khodjamirian et al., {\sl Phys. Lett.} {\bf B457}, 25 
(1999).

\bibitem{gro} A.G. Grozin and O.I. Yakovlev, {\sl Eur. Phys. J.} {\bf C2},
721 (1998).

\bibitem{dn} H.G. Dosch and S. Narison, {\sl Phys. Lett.} {\bf B368}, 163 
(1996).

\bibitem{ddpi} F.S. Navarra et al., {\sl Phys. Lett.} {\bf B489}, 319 
(2000). 

\bibitem{rry}  L.J. Reinders, H. Rubinstein and S. Yazaki, {\sl Phys.
               Rep.} {\bf 127}, 1 (1985). 

\bibitem{ddrho} M.E. Bracco et al., hep-ph/0108223.

\bibitem{io2}  B.L. Ioffe and A.V. Smilga, {\sl Nucl. Phys.} {\bf B216} 373
(1983); {\sl Phys. Lett.} {\bf B114}, 353 (1982).

\bibitem{cut} R.E. Cutkosky, {\sl J. Math Phys.} {\bf1}, 429 (1960).

\bibitem{svz}  M.A. Shifman, A.I. Vainshtein and  V.I. Zakharov, {\sl Nucl. 
Phys.} {\bf B147} 448 (1979).

\bibitem{BBG93} E. Bagan, P. Ball and P. Gosdzinsky, Phys. Lett {\bf B301}, 
249 (1993).

\bibitem{ra} R. S. Marques de Carvalho et al., Phys. Rev. {\bf D60}, 
034009 (2000);H.G. Dosch et al., Phys. Lett. {\bf B431}, 173 (1998).

\bibitem{io1} B.L. Ioffe and A.V. Smilga, {\sl Nucl. Phys.} {\bf B232},  109
(1984).

\bibitem{dnn} F.O. Dur\~aes, F.S. Navarra and M. Nielsen, {\sl Phys. Lett.}  
             {\bf B498},  169  (2001). 

\bibitem{cas} R. Casalbuoni et al., {\sl Phys. Rep.} {\bf 281}, 145 (1997).

\bibitem{sin} P. Singer, {\sl Acta Phys. Polon.} {\bf B30}, 3849 (1999).


\end{thebibliography}
\end{document}